\documentclass[twocolumn,aps,amsmath,showpacs,pre]{revtex4}
\usepackage{graphicx}
\usepackage{amssymb}
\usepackage{color}

\begin{document}
\title{Entropy of rigid $k$-mers on a square lattice}
\author{Lucas R. Rodrigues}
\email{lucasr@id.uff.br}
\affiliation{Instituto de F{\'{\i}}sica, Universidade Federal Fluminense,
  Niter\'{o}i, Brazil}
\author{J.F. Stilck}
\email{jstilck@id.uff.br}
\affiliation{Instituto de F{\'{\i}}sica and National Institute of Science
  and Technology for Complex Systems, Universidade Federal Fluminense,
  Niter\'{o}i, Brazil} 
\author{W. G. Dantas}
\email{wgdantas@id.uff.br}
\affiliation{Departamento de Ci\^{e}ncias Exatas, EEIMVR,
  Universidade Federal Fluminense,
  Brazil} 
\date{\today}

\begin{abstract}
  Using the transfer matrix technique, we estimate the entropy for a gas
  of rods of sizes equal to $k$ (named $k$-mers), which cover completely
  a square lattice. Our calculations were made considering three different
  constructions, using periodical and helical boundary conditions. One
  of those constructions, which we call {\it{Profile Method}}, was based
  on the calculations performed by Dhar and Rajesh \cite{dr21} to obtain
  a lower limit to the entropy of very large chains placed on the square
  lattice. This method, so far as we know, was never used before to define
  the transfer matrix, but turned out to be very useful, since it produces
  matrices with smaller dimensions than those obtained using other approaches.
  Our results were obtained for chain sizes ranging from $k=2$ to $k=10$
  and they are compared with results already available in the
  literature. In the case of dimers ($k=2$) our results are compatible with
  the exact result, for trimers ($k=3$), recently investigated by Ghosh
  {\it{et. al}} \cite{ghosh} also our results were compatible, the same
  happening for the simulational estimates obtained by
  Pasinetti {\it{et. al.}} \cite{pasinetti} in the whole range of rod sizes.
  Our results are consistent with the asymptotic expression
  for the behavior of the entropy as a function of the size $k$, proposed
  by Dhar and Rajesh \cite{dr21} for very large rods ($k \gg 1$).
\end{abstract}


\maketitle

\section{Introduction}
\label{intro}

We study a system of rigid rods formed
by $k$ consecutive monomers placed on the square lattice.
This is a problem which has a long history in statistical mechanics.
The particular case when $k=2$ (dimers) in the full lattice limit, when the
rods occupy all sites of the lattice, is one of the few exact solutions
of interacting models which were obtained so far \cite{ktf61}.
Another aspect of the thermodynamic behavior of long
rod-like molecules was already anticipated by Onsager in the 40's: he
argued that at high densities they should show orientational (nematic)
order \cite{o49}, due to the excluded volume interactions. In a seminal paper,
for the case of rods on the square lattice, \cite{gd07} Ghosh and Dhar found,
using simulations, that for $k \ge 7$ at low density of rods an isotropic
phase appears, but as the density is increased a continuous transition to
a nematic phase happens. Evidence was found that close to the
full lattice limit the orientational order disappears at a density
$1-\rho_c \sim k^{-2}$. The presence of the
nematic phase at intermediate densities of rods was proven rigorously
\cite{dj13}. Because simulations at high densities of rods are difficult, an
alternative simulational procedure allowed for more precise
results for the transition from the nematic to the the high density
isotropic phase \cite{k13}. Recent results suggested this transition
to be discontinuous \cite{sdr22}.

Here we consider the estimation of the entropy of $k$-mers on the square
lattice in the full lattice limit, for $k \ge 2$. This has been discussed
before in the literature. Baumg\"{a}rtner \cite{b85} generated exact
enumerations of rods for $2 \le k \le 12$ on $L \times L$ square lattices,
but did not attempt
to extrapolate his results to the two-dimensional limit $L \to \infty$. His
interest was actually more focused on the question if the system is isotropic
or nematic in this limit. Bawendi and Freed \cite{bf86}
used cluster expansions in the
inverse of the coordination number of the lattice to improve on mean field
approximations. For dimers on the square lattice, their result is about
8 \% lower than the exact result \cite{ktf61}, and there are indications
that the differences are larger for increasing rod lengths $k$. A study
of trimers ($k=3$) on the square lattice using transfer matrix techniques
similar to the ones we use here, was undertaken by Ghosh, Dhar and
Jacobsen \cite{ghosh} and has led to a rather precise estimate for the
entropy. Computer simulations have also been useful in this field, and
estimates for the entropy of $k$-mers on the square lattice were obtained
by Pasinetti et al \cite{pasinetti} for $2\le k \le 10$, besides studying
other statistical properties of the high density phase of the system. Another
analytic approximation to this problem may be found in the paper by
Rodrigues, Stilck and Oliveira \cite{rso22}, where the solution of the
problem of rods on the Bethe lattice for arbitrary density of rods \cite{drs11}
was performed for a generalization of this lattice called the Husimi lattice.
These solutions on the central region of treelike lattices may be seen as
improvements of mean field approximations to the problem. Again there are
evidences that the quality of the estimates decreases for increasing values
of $k$, while the difference of the estimate for dimers to the exact value
is of only 0.03 \%, it already grows to 3 \% when compared to the estimate
for trimers presented in \cite{ghosh}.

The approach we employ here to study the problem is to formulate it in terms
of a transfer matrix, as was done for trimers in \cite{ghosh}. It consists to
define the problem on strips of finite widths $L$ with periodic and helical
boundary conditions in the finite transverse direction. The leading
eigenvalue of the transfer matrix determines the entropy of the system,
as will be discussed below. The values of the entropies for growing widths
are then extrapolated to the two-dimensional limit $L \to \infty$, generating
estimates and confidence intervals for each case. For the case of periodic
boundary conditions (pbc), besides using the conventional
definition of the transfer
matrix, in which $L$ sites are added at each application of it, we used an
alternative approach, inspired on the generating function formalism which
was developed by Dhar and Rajesh in \cite{dr21} to obtain a lower bound for 
the entropy of the system. This alternative procedure turned out to be more
efficient for this problem than the conventional one, in the sense that the
size of the transfer matrices were smaller, thus allowing us to solve the
problem for larger widths $L$. For helical boundary conditions (hbc), only the
conventional formulation of the transfer matrix was used.

Finally, we already mentioned that the possible orientational ordering of
the rods in the full lattice limit was, for example, a point which motivated
the exact enumerations in \cite{b85}. For dimers, it is known exactly that
no orientational order exists \cite{hl72}, but on the square and hexagonal
lattices, which are bipartite,  orientational correlations decay with a
power law \cite{ktf61}, while there is no long range orientational order
on the triangular lattice in the same limit \cite{f02}. This point is also
investigated numerically for trimers in \cite{ghosh}, with compelling
evidences that the dense phase in the full lattice limit is not only critical
but has conformal invariance. As already mentioned, so far all indications are
that the high density phase of the system is isotropic on the square
lattice, possibly with orientational correlations decaying with a power law.

This paper is organized as follows: the construction of the transfer
matrices and determination of the leading eigenvalues and the entropies
are described in section \ref{TM}. The numerical results for the entropies
of the rods on strips, the extrapolation procedure and the estimates
for the entropy of the rods on the square lattice may be found in section
\ref{NR}. Final discussions and the conclusion are found in section \ref{conc}.
      
\section{Transfer matrix, leading eigenvalues and entropy}
\label{TM}

The transfer matrix will be determined by the approach used to describe the
transverse configurations of the strips at different levels, which
define the states of
lines and columns of the matrix. The two approaches we used are described
below. We consider a lattice in the $(x,y)$ plane, with
$1\leq x\leq L$ and $0\leq y\leq\infty$. Periodical or helical boundary
conditions are used in the transverse direction, that is, horizontal
bonds are placed between sites $(L,y)$ and $(1,y)$ in the first case
and $(L,y)$ and $(1,y+1)$ in the second case. Fixed boundary conditions
are used in the longitudinal direction.
For periodic boundary conditions, we use two approaches in order to
obtain the transfer matrix, which we call {\it{Usual Approach}}
and the {\it{Profile Method}}. Those two approaches, although defining
the states of the matrix in different ways, will of course produce
exactly the same
results for the entropy per site of chains with length  $k$ placed on
a lattice with width  $L$. For helical boundary conditions, only the
{\it{Usual Approach}} is used. In the following, we shall describe each of
those approaches.

\subsection{Periodical Boundary Conditions}
In the {\it{Usual Approach}}, at each application of the
transfer matrix $L$ new sites are incorporated into the lattice, while
in the second approach a variable number of $k$-mers is added to the system
at each step, so that the ensemble is grand-canonical in this case.

\subsubsection{Usual Approach}
This way to build the transfer matrix is the same
used by Ghosh {\it{et. al.}} \cite{ghosh} (named as ``Second Construction'')
to study the case of trimers $(k=3)$. It was also applied in
previous works by two of the authors \cite{dantas1,dantas2}.
As mentioned in those
papers, this method is inspired by the work of Derrida \cite{derrida}, which
applied it to  the problem of an infinite chain placed on a
cylinder.

\begin{figure}[h!]
\begin{center}
  \includegraphics[scale=0.5]{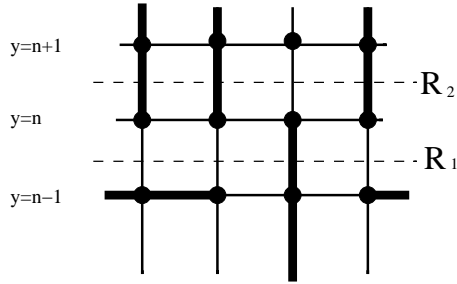}
  \caption{Example of a possible continuation for a state defined by vertical
    bonds in a stripe of width $L=4$, identified by the reference
    line $R_1$, followed by a state connected to it defined by the
    reference line $R_2$.}
    \label{fig1}
\end{center}
\end{figure}

The states which define the transfer matrix in this formulation are
determined by the possible configurations of the set of $L$ vertical
lattice edges cut by a horizontal reference line which is located between
two rows of horizontal edges of the lattice, such as the dashed lines $R_1$ and
$R_2$ in Fig. \ref{fig1}. These states may be represented by a vector,
where each component corresponds to the number of monomers already connected
to it, i.e, those located on sites below the reference line. Thus,
the components are restricted to the domain $[0,k-1]$. So, from
the information given by this vector, we can find all possible configurations
of the vertical edges cut by the reference line situated one lattice spacing
above, allowing us to define the transfer matrix of the problem. An illustration
of possible configurations and their representative vectors for pbc can
be observed in figure \ref{fig1}, where we have a state defined for the
case $k=3$ and $L=4$. At the reference line $R_1$, separating the levels
$y_{n-1}$ and $y_n$, we have the vector $|v_1\rangle=(0,0,2,0)$, while at
$R_2$, linking the levels $y_n$ and $y_{n+1}$, the configuration is represented
by $|v_2\rangle=(1,1,0,1)$.

We proceed developing an algorithm to obtain the elements
of the transfer matrix, for given values of $k$ and $L$. However, we are
limited by the amount of computational memory and by the time necessary
to compute those elements. For a given value of $k$, the number of states
grows roughly exponentially with $L$. Even considering rotation symmetry, which
makes states such $|v_1\rangle=(0,1,0)$ and $|v_2\rangle=(0,0,1)$ equivalent,
and reflection symmetry, where $|v_1\rangle = (0,1,2,3)$
and $|v_2\rangle=(3,2,1,0)$ can be treated as the same state,
this property imposes an upper limit to
the widths that we are able to study for each rod size, $k$.

In principle, without considering the reduction of the size of the transfer
matrix due to symmetries, one would suppose that this size would be equal
to $k^L$, but the transfer matrix is actually block diagonal, each state
being associated to one of the blocks. It happens that the leading eigenvalue
always belongs to the block generated by the state
$|v_0\rangle=(0,0,\ldots,0)$. So, instead of determining the entire transfer
matrix, we proceed using the same strategy developed by Ghosh {\it{et. al.}}
\cite{ghosh} for trimers, generating the subset of states which starts
with the state $|v_0\rangle$ and generating all other states connected to it.

Once we compute the transfer matrix, $\mathcal{T}$, to obtain the value of the
entropy per site for the case of gas of monodisperse rigid chains with size
$k$ in a strip of size $L$, we may then compute the dimensionless entropy
per lattice site
\begin{eqnarray}
  s_L=\lim_{N\to\infty}\frac{S}{Nk_B}=\lim_{N\to\infty}\ln\Omega,
\end{eqnarray}
where $N=L\ell$ is the number of the sites and $\Omega$ is the number
of configurations of the rods of size $k$ placed on the strip. So, that
number is related with the transfer matrix as $\Omega=Tr(\mathcal{T}^\ell)$ and
if $\lambda_1$ is the largest eigenvalue of $\mathcal{T}$, we get, in the
thermodynamic limit $\ell \to \infty$:

\begin{eqnarray}
  s_L=\frac{1}{L}\ln\lambda_1.
\end{eqnarray}

So, to obtain the entropy of a given width $L$, we should determine
the largest eigenvalue for the transfer matrix. Fortunately, the
typical transfer matrix is always very sparse,
which allows us to use a method such as the Power Method, so that the
determination of this eigenvalue becomes a possible task for quite large
widths.

\subsubsection{Profile Method}
\begin{figure}[h!]
\begin{center}
  \includegraphics[scale=0.7]{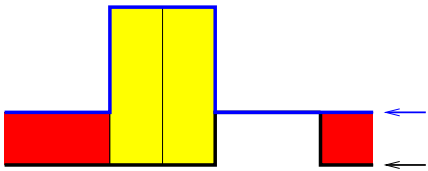}
  \caption{Illustration of one step of the process of filling the stripe
    of width $L=7$ with trimers ($k=3$). The initial profile is the thick
    black line and its baseline is at the level pointed by the black arrow.
    The height profile in this case will be $(0,0,0,0,1,1,0)$, notice that
    there are two steps (5 and 6, from left to right) which are at the
    same height in both profiles. One possibility is to aggregate one
    horizontal rod (red on line) and two vertical rods (yellow on line).
    The new baseline is pointed at by the blue arrow and the new profile will
    be represented by $(0,0,2,2,0,0,0)$. The contribution of this
    configuration is $z^3$.}
    \label{prof}
\end{center}
\end{figure}

This alternative method of defining the transfer matrix is inspired on
the generation function approach used by D. Dhar and R. Rajesh to obtain
lower bounds for the entropy of k-mers in the full lattice limit \cite{dr21}.
It is convenient in this case to consider the dual square lattice, whose center
of elementary squares correspond to sites in the lattice of the previous
section, and represent the $k$-mers as $k \times 1$ rectangles on this
lattice. Unlike the {\it{Usual Approach}}, where $L$ sites are added at
each multiplication
of the transfer matrix, in this method a variable number of $k$-mers is
added at each step. We consider the profile of the upper end of the
stripe at a particular
point in filling it up with rods, such as the one shown in Fig. \ref{prof},
as defining the states to build the transfer matrix.
For a particular profile, we define the {\em baseline} as the horizontal
line passing through the lowest points of the profile. We then consider
the operation of adding rods to all points in the baseline, so that a new
baseline is generated at a level at least one lattice parameter higher than
the previous one. There may be more than one way to accomplish this,
involving different numbers of added rods. We will denote by $z$ the
fugacity of one rod, so that the contribution to the element of the
transfer matrix corresponding to a particular choice of new rods added to
the stripe will be $z^{n_r}$, where $n_r$ is the number of new rods added
to the stripe. Notice that no $k$-mer will be added which will not have
at least one monomer located on the baseline. The profiles, which define
the states of the transfer matrix,
may be represented by a vector with $L$ integer components, ranging
between $0$ (the intervals on the baseline) and $k-1$. Thus, in general,
there will be $k^L$ possible states. However, as mentioned before,
this general transfer matrix
will be block diagonal, and as was done before for the case of
trimers \cite{ghosh} we will restrict ourselves to the subset of states
which include the horizontal profile $(0,0,\ldots,0)$, since in all cases
where  we were able to consider all profiles the leading eigenvalue of
the transfer matrix was found in this block. 

In the grand-canonical ensemble we are considering, let $M$ be the number
of times the transfer matrix is applied. In the thermodynamic limit
$M \to \infty$ the partition function $Y_M(z)$ will be determined by
the leading eigenvalue $\lambda_1$ of the transfer matrix
$Y_M(z) \approx \lambda_1^M$, so that the thermodynamic potential will be:
\begin{equation}
  \Phi(T,V,\mu)=-k_BT \ln Y_M(z)=-k_BTM \ln \lambda_1(z),
\end{equation}
where $z=e^{\beta \mu}$, $\mu$ being the chemical potential of a rod. 
The entropy will be given by the state equation 
\begin{eqnarray}
  S(z)&=&-\left(\frac{\partial \Phi}{\partial T}\right)_{M,\mu}= \nonumber\\
  &&k_BM\left[ \ln \lambda_1(z)
    -z\ln z \frac{1}{\lambda_1(z)}
    \frac{\partial \lambda_1(z)}{\partial z}\right],
\end{eqnarray}
and the total number of rods will be
\begin{eqnarray}
  N_r(z)&=&-\left(\frac{\partial \Phi}{\partial \mu}\right)_{T,M}= \nonumber \\
  &&M\frac{z}{\lambda_1}\left(\frac{\partial \lambda_1}{\partial z}\right).
\end{eqnarray}
The dimensionless entropy per lattice site occupied by rods will then be:
\begin{equation}
s(z)=\frac{S(z)}{k_BN_r(z)k}=\frac{\ln \lambda_1(z)}
{\frac{kz}{\lambda_1}\left(\frac{\partial \lambda_1}{\partial z}\right)}-
\frac{\ln z}{k}.
\label{sz}
\end{equation}
In the grand-canonical ensemble, the remaining extensive
variable of the potential is usually
the volume. The number of rods will be different in the configurations
which contribute to the partition function, and by construction they occupy
the lower part of the lattice in a compact way. For simplicity, let us consider
widths $L$ that are multiples of $k$. We then see that for a given value of M,
the height $H$ of the region occupied by the rods will be in the
range $[M,kM]$, so that the volume should be at least equal to $L
\times kM$. Actually, it could be fixed at any value above this one without
changing the results. This means that this condensed phase of
$k$-mers actually coexists with the part of the lattice which is empty, and
since the grand canonical potential of the coexisting phases should be equal
we conclude that $\Phi(T,V,\mu)=0$, because this will be the potential of
the phase which corresponds to the empty lattice. In other words,
we recall that the grand canonical potential is proportional to the
pressure (force per unit length in the two-dimensional case), which
should be the same in the coexisting phases. This condition of coexistence
determines the activity of a rod
\begin{equation}
  \lambda_1(z_c)=1,
  \label{zc}
\end{equation}
and substitution of this restriction into Eq. \ref{sz} leads to the final
result for the entropy per site occupied by the rods in this formulation
of the transfer matrix:
\begin{equation}
  s_L=-\frac{\ln z_c}{k}.
  \label{els}
\end{equation}
In summary, in the formulation where the states are determined by the height
profile of the $k$-mers in the stripe, we solve numerically Eq. \ref{zc} for
the activity $z_c$ which corresponds to a vanishing pressure of the
condensed phase of rods and then determine the entropy per site of this phase
using Eq. \ref{els}.
 
It is then interesting to consider explicitly the simplest non-trivial
case using the {\it{Profile Method}},
which is $L=k$. Starting with the horizontal profile, we notice that for
$L=k$ there will be two possibilities to add a new set of rods and shift
the baseline upwards: either a single horizontal rod or $k$ vertical rods
are added, and the new profile is again horizontal in both cases. Due to
the periodic boundary conditions, in the first case there are $L=k$ different
ways to place the horizontal rod. We thus
conclude that there is a single profile state in this case and the size
of the transfer matrix is $1 \times 1$ so that
\begin{equation}
  \lambda_1=kz+z^k.
\end{equation}
We see then that $z_c$ is defined by the equation $z_c^k+kz_c-1=0$ and the
entropy per site will be given by Eq. \ref{els}.

\begin{figure}[h!]
\begin{center}
  \includegraphics[scale=0.75]{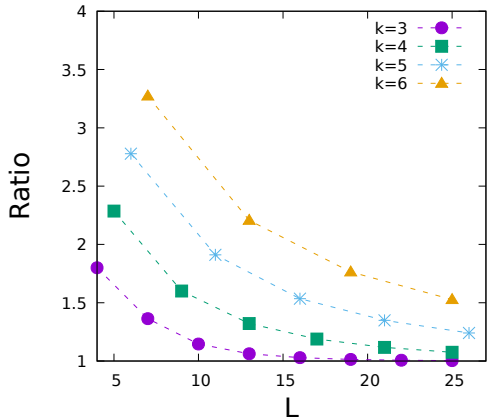}
  \caption{Ratio between the number of states of the transfer matrix
    considering the {\it{Usual Approach}} and the {\it{Profile Method}}.
    The dashed lines are just a guide for the eye.}
  \label{fig3}
\end{center}
\end{figure}

We proceeded using both methods described above to calculate the entropy
for a set of rod sizes $k$ and growing widths $L$. To reduce the size of
the transfer matrices, we use rotational and reflection symmetries of the
states. Since we want to obtain estimates for the entropies per site in
the two-dimensional limit $L \to \infty$, it is important to reach the
largest possible widths $L$ for each rod size $k$. It should be noticed
that in the {\it{Profile Method}} we have to solve numerically Eq. \ref{zc} for
$z_c$, so that the leading eigenvalue $\lambda_1$ has to be calculated
several times, while in the conventional method  only one determination
of the leading eigenvalue is needed. This seems to indicate that the
{\it{Usual Approach}} should allow
us to reach the largest widths. However, if we compare the numbers of
states (size of the transfer matrices) in both methods, we obtain the
results shown in Fig. \ref{fig3}. We notice that the transfer matrices
are systematically larger for the {\it{Usual Approach}}, the difference
increasing
monotonically with the rod size. Therefore, at the end the {\it{Profile Method}}
allowed us to reach the largest widths in all cases, which are determined
by the limitations in time and memory of the computational resources
available to us. 

\subsection{Helical boundary conditions}
\begin{figure}[h!]
\begin{center}
  \includegraphics[scale=0.7]{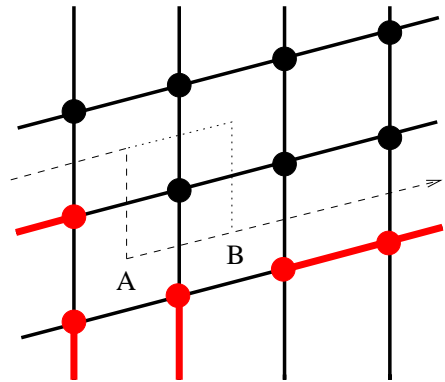}
  \caption{Stripe of width $L=4$ with helical boundary conditions in
    the transverse direction. The $L+1$ lattice edges crossed by the
    dashed line, starting at point A and in the direction indicated by
    the arrow define the vector which represents the state at this point.
    An additional site is incorporated when the transfer matrix is
    applied, so that the new starting point of the line is B. Trimers
    are represented by thick lines (red on line).}
  \label{hbc}
\end{center}
\end{figure}

An alternative way to define the boundary conditions of stripes of width
$L$ is to make them helical. This was already used by Kramers and Wannier
in their seminal paper about the Ising model \cite{kw41}. To visualize
these boundary conditions, if we consider the model on a cylinder with
perimeter of size $L$, the transverse lattice edges are on a helix with
pitch $L$ as seen in Fig. \ref{hbc}. The states are defined, as in the
{\it{Usual Approach}}, by the number of monomers already incorporated
into the rods on the
$L+1$ edges cut by a line which divides the stripe into two sectors.
In the {\it{Usual Approach}}, this line, as may be seen in Fig. \ref{fig1},
is horizontal
and cuts $L$ edges, while for helical boundary conditions it is also
parallel to the transverse edges for $L$ steps, ending with a vertical
step. This is illustrated by the dashed line in Fig. \ref{hbc}, at a
given step the line starts at point A, cuts $L$ vertical edges and
finally cuts an additional transverse edge. All sites below the curve
are occupied by monomers. While for periodic boundary conditions $L$
lattices sites are added to the system as the transfer matrix is applied
(the sites between lines $R_1$ and $R_2$ in Fig. \ref{fig1}), a single
site is added for helical boundary conditions, the line which defines
the new state starts at point B and the last two steps of the previous
line are replaced by the dotted steps. For the particular case in the
figure, the state associated to the line starting at A is $(0,0,0,0,0)$,
while there are two possibilities for the state B: $(0,0,0,0,1)$ or
$(0,0,0,1,0)$, since a new trimer has to start at the edge incorporated
in this step and it may be horizontal or vertical. 

Thus, an important aspect of this boundary condition, as compared to
the periodic one, is that only one or two elements of each line of
the transfer matrix are equal to 1, all others vanish, so in general
they lead to sparser transfer matrices, which of course is desirable
if we use the Power Method to calculate the leading eigenvalue. The
drawback is that the reflection and rotation symmetries are not
present in this case.  

\section{Numerical Results}
\label{NR}

In this section, we discuss the numerical results obtained using the three
approaches to determine the transfer matrix for the case of a
monodisperse gas of rigid chains with size $k$,
filling a strip of width $L$ with periodical and helical boundary conditions.

Besides presenting the values of the entropy for each case, we also
discuss the question of the transfer matrix dimension, which turns
out to be the major obstacle in obtaining the entropy for a given
$(k, L)$ pair. Also, after collecting some figures for the entropies
we should deal with the task of how to extrapolate them to obtain
an estimate for $s_{\infty}(k)$, from a set $\{s_L(k)\}$, to the
two-dimensional limit, i.e, when $L\to\infty$. For that, we are aware that
for critical two-dimensional isotropic statistical systems, presenting only
short-range interactions, conformal invariance predicts that in a
cylinder of width $L$, the entropy per site must follow the asymptotic
behavior \cite{c87},
\begin{eqnarray}
  s_L(k)=s_{\infty}(k)+\frac{A}{L^2}+o(L^{-2}),
  \label{eq3}
\end{eqnarray}
where $A$ is related to the central charge.

Using the methods previously described to build the transfer matrix,
we could determine the entropy of a given size of rods $k$ for different values
of $L$, limited by the amount of computer memory required
in each case and/or by the processing time. Once we have obtained the elements
of the matrix, the calculation of its dominant eigenvalue was carried out
using the Power Method to diagonalize the matrix,

\begin{eqnarray}
  \mathcal{T'}=\mathcal{T}+p\mathcal{I},
\end{eqnarray}
where $p$ is a positive real number and $\mathcal{I}$ is the unitary matrix.
Such procedure was necessary because the original matrix, $\mathcal{T}$, usually
has a set of dominant eigenvalues, which, despite always presenting
at least one of those
eigenvalues at the real axis, has others with the same modulus in the
complex plane. Such feature turns out to be a difficulty for the
Power Method to
work properly. However, using that translation we can shift the eigenvalues
along the real axis all the eigenvalues, making the positive
real one the only dominant
eigenvalue, $\lambda'$ for the matrix $\mathcal{T'}$. Then, to recover the
value which we are looking for, $\lambda$, we have
$\lambda=\lambda'-p$.

The choice of the parameter $p$ may be a sensitive issue in order to
get the right results for the dominant eigenvalues in an efficient way.
We adopted the strategy to fix this parameter maximizing 
the ratio between the real positive eigenvalue and the one with the the second
largest modulus. However, in the cases
we verified here, even spanning the values of $p$ over a large interval,
such is $[1:100]$, only minor differences among the results $(\approx 10^{-14})$
appear. In fact, 
the only noticeable effect caused by changing the size of this translation
is observed in the number of steps needed to the Power Method converge
with a given precision (in our case this precision is about $10^{-13}$).
For growing values of $p$ the number of steps increases, roughly, in a
linear fashion.

\begin{figure}[h!]
\begin{center}
  \includegraphics[scale=0.75]{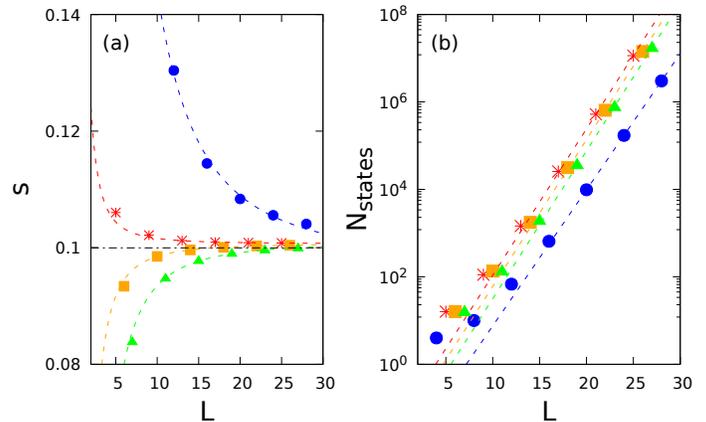}
  \caption{Panel (a): Behavior of the entropy for tetramers ($k=4$) as
    a function of $L$ separated
    by sets, where dots, stars, squares, and triangles are related to the rests
    R=0,1,2 and 3, respectively. The dashed lines are fittings of each
    set according to the relation given by Eq. \ref{eq3}. Panel (b) shows
    how the number of states of
    the transfer matrix grows as a function of L for each of the sets. The
    dashed lines here are used just as a guide for the eyes, indicating the
    exponential behavior of that number when $L$ is large enough.}
    \label{fig2}
\end{center}
\end{figure}

Just as it happens for trimers \cite{ghosh}, each other $k$-mer has its
entropy values following the relation Eq. \ref{eq3} in separate sets
depending on the rest, $R$, of the division $L/k$. Hence, if for trimers we
have three sets (values with remainders 0, 1, and 2), in other cases
there will be $k$ sets of values for the entropy
obeying the asymptotic behavior, as we can see in the figure \ref{fig2}(a)
for the case $k=4$. Such behavior obviously poses an additional
difficulty in order to get from each set a good extrapolation for
the entropy in the two-dimensional limit, when $L\to\infty$.

Now we start to discuss the results obtained from each of the approaches
presented in the previous section, considering its peculiarities and the
limitations of each of them concerning widths which could be reached.

\subsection{Periodic boundary conditions}
For these boundary conditions, we applied the {\it{Usual Approach}} and the
{\it{Profile method}}. As already mentioned before, the
{\it{Profile Method}} turns
out to be more effective for larger values of $k$ and $L$. We will thus
restrict ourselves to present the results furnished by that method,
after remarking that we have verified that for trimers the numbers of
states we have obtained using the {\it{Usual Approach}} are equal to the ones in
reference \cite{ghosh} obtained from the second construction. Of course,
as already mentioned before, both approaches lead to the same values for
the entropies.

The dimension of the transfer matrix, for a given
value of $k$, grows nearly exponentially as a function of $L$ - as
we can see in figure \ref{fig1}(b) - considering
the behavior for each set of values of a given remainder R.
Then, for a high value of $k$ the number of elements for the set $\{s_L\}$ 
cannot be as large as it is when we consider smaller chains.
The dimension reached in our calculations using the {\it{Profile Method}},
for each set of remainder R is discriminated in table \ref{tab1}. 

\begin{table}[h!]
\begin{center}
  \begin{tabular}{|c||c|c|c|c|c|c|c|c|c|c|}
    \hline
    k&$R_0$&$R_1$&$R_2$&$R_3$&$R_4$&$R_5$&$R_6$&$R_7$& $R_8$& $R_9$\\
    \hline
    \hline
    2& 13 &13& & & & & & & & \\
    \hline
    3& 11 &10&10& & & & & & & \\
    \hline
    4& 8 &7&7&7& & & & & & \\
    \hline
    5& 7 &6&6&6&6& & & & & \\
    \hline
    6&7&6&6&6&6&6& & & & \\
    \hline
    7&6&5&5&5&5&5&5& & & \\
    \hline
    8&6&5&5&5&5&5&5&5& & \\
    \hline
    9&5&4&4&4&4&4&4&4&4& \\
    \hline
    10&4&4&4&4&4&4&4&4&4&3 \\
    \hline
  \end{tabular}
  \caption{Number of elements for the set $\{s_L\}$ for different
    sizes of the chains. Each set $R_i$ is related to the remainder,
    $i=0,1,2,...$ for the division $L/k$.}
    \label{tab1}
\end{center}
\end{table}

Using the entropy values in each set, we can obtain an extrapolated
result for $s_{\infty}(k)$. This was done using the approach known
as BST extrapolation method \cite{henkel}. Since this method can be
functional even in situations where the number of entries to extrapolate
is not that big, it appears to be convenient to use it in our problem.
As is described in the reference \cite{henkel}, the BST method has
a parameter $\omega$, which
in our case should be set as $\omega=2$, due to the relation Eq.\ref{eq3}.
Also, because the desired limit, $s_{\infty}$, is obtained from a table
of extrapolants, $T_m^{(i)}$, where $m$ is related to the extrapolant generation,
then the uncertainty of the estimate will be defined as:

\begin{eqnarray}
  \sigma=2|T_m^{(1)}-T_m^{(0)}|,
  \label{eq5}
\end{eqnarray}
when $m\to\infty$. In practical terms, this limit is applied
considering the difference between the two approximants before the
last generation. As an example, if we get $N$ entries, then the
extrapolated value is obtained from the $(N-1)$'th generation,
while the uncertainty is calculated from the two approximants of the
$(N-2)$'th generation.

\begin{table}[h!]
\begin{center}
  \begin{tabular}{|c|c|c|c|c|c|}
    \hline
    $k$&$s_i(\sigma_i)$& $k$& $s_i(\sigma_i)$&$k$& $s_i(\sigma_i)$\\
    \hline
    2& 0.2915609067(66)  &3 &0.1584937(64) &4&0.100669(73)\\
    &0.29156090403(14)   &  &0.15850495(19)& &0.1007572(55) \\
    &                    &  &0.158510(25)  & &0.1007747(48)\\
    &                    &  &              & &0.100780(87) \\
    \hline
   5&0.0700(33)      &6 &0.054(29)     &7& 0(2)\\
    &0.0703370(82)   &  &0.05210(76)   & &0.04030(68)\\
    &0.070350(44)    &  &0.0522275(66) & &0.040475(85)\\
    &0.07038344(58)  &  &0.052244(84)  & &0.0404471(96)\\
    &0.070303(71)    &  &0.05228(15)   & &0.040530(11)\\
    &                &  &0.052193(66)  & &0.040548(16)\\
    &                &  &              & &0.040561(44)\\
    \hline
    8&0.0164(23)       &9 &0.016(14)     &10&0.015(14)\\
     & 0.03243(22)     &  &0.02664(25)   &  &0.02223(23)\\
     & 0.03243(16)     &  &0.02567(24)   &  &0.02226(27)\\
     & 0.03243(11)     &  &0.02660(43)   &  &0.02229(31)\\
     & 0.0324476(43)   &  &0.026633(29)  &  &0.022316(57)\\
     & 0.032466(11)    &  &0.026704(73)  &  &0.022337(12)\\
     & 0.032483(22)    &  &0.0267322(91) &  &0.022355(55)\\
     & 0.032499(38)    &  &0.026752(67)  &  &0.02237(11)\\
     &                 &  &0.02677(14)   &  &0.02238(16)\\
     &                 &  &              &  &0.22400(21)\\
    \hline
  \end{tabular}
  \caption{Extrapolated entropy values for each set of a given chain size
    $k$. Results for periodic boundary conditions. For each value of $k$,
    the extrapolated entropies and uncertainties for remainders
    $R=0,1,\ldots,k-1$ are presented. Those values and its uncertainties
    were obtained using the BST method with $\omega=2$ and with the uncertainty
    determined by the Eq. \ref{eq5}.}
    \label{tab2}
\end{center}
\end{table}

So, using the BST extrapolation method we were able to obtain the values
shown in table \ref{tab2} for each set associated with the remainder
of the ratio
$L/k$. To finally get a value $\bar{s}_{\infty}(k)$, representing the
extrapolation for all sets considered,  we calculate an average
and a total uncertainty
weighted by the uncertainties of each value of $s_i$, obtained for
some remainder $R$.
Once we consider the values $s_i(k)$ statistically independent of each other,
the average and its deviance have to obey the following relations,

\begin{eqnarray}
  \overline{s}_{\infty}&=&\frac{\sum_i s_i/\sigma_i^2}{\sum_i 1/\sigma_i^2
  }\nonumber\\
  \Delta s_{\infty}&=&\sqrt{\frac{1}{\sum_i 1/\sigma_i^2}},
  \label{eq12}
\end{eqnarray}
where $s_i$ is the extrapolated value of the entropy for some set of ratio R,
while $\sigma_i$ is the uncertainty related to it, which is obtained from the
equation Eq. \ref{eq5}. The results of the final values of the entropies
of each size $k$ are shown in the table \ref{tab5}, which also shows, for
comparison, the
correspondent values obtained from Monte Carlo numerical simulations
developed by Pasinetti {\it{et. al.}} \cite{pasinetti}.

Notice from the table \ref{tab2} that as the size $k$ of the rods grows,
the precision
for the values $s_i$ is smaller, since the number of entries for each set
$\{s(k)\}$ diminishes. It is also perceivable that the sets associated with
the remainder $R=0$ lead to the worst results for the extrapolation.
This happens because such cases have a slower approach to the limit
$\overline{s}_{\infty}(k)$. Therefore, that set, although related to the smaller
transfer matrix dimensions, needs a larger number of entries to produce a better
result. On the other hand, our final results are in excellent agreement with the
exact value obtained for the dimer case $(k=2)$ \cite{ktf61}, where
$s_{\infty}(2)=G/\pi$, where $G\approx 0.9159655941772...$ is 
Catalan's constant. From the table \ref{tab5} we see that our estimate
coincides with this exact result up to the 11th decimal place. Also, for
the case $k=3$, we can compare our result with that obtained by Ghosh
{\it{et. al}} \cite{ghosh}, i.e,
$\overline{s}_{\infty}(3)=0.158520(15)$, which is also in accordance with the one
shown in table \ref{tab5}. For the rest of the cases, we have also a good
agreement with the results obtained by Pasinetti {\it{et. al}}
\cite{pasinetti} through numerical simulations, although, our
extrapolations exceed their values, in precision, at least in one order
of magnitude.

\begin{figure}[h!]
\begin{center}
  \includegraphics[scale=0.75]{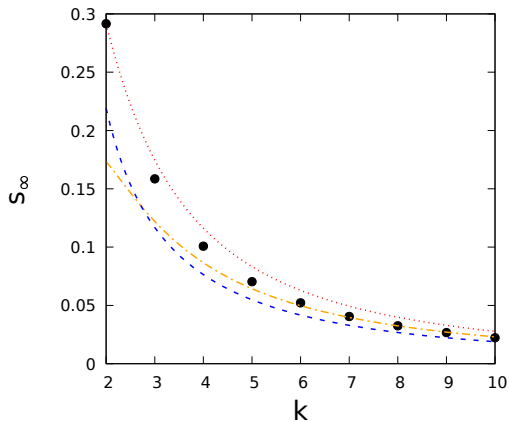}
  \caption{Extrapolated entropies of rods filling a square lattice
    as a function of the size of chains, $k$. The blue dashed
    and the red dotted lines correspond, respectively, to the lower and
    upper bounds, obtained in \cite{dr21} and \cite{gagu}(Eqs. \ref{lb}
    and \ref{ub}, respectively).
    The dashed-dotted line follows the behavior predicted
    by Dhar and Rajesh \cite{dr21} when $k\to\infty$, i.e, $s=\ln k/k^2$.}
    \label{fig4}
\end{center}
\end{figure}

In figure \ref{fig4} we can see how the entropy globally behaves
as a function of the chain size $k$. First of all, such values are
constrained between two limits. A lower bound,

\begin{eqnarray}
  s(k)\geq s_{2k\times\infty}\geq \frac{\ln k}{k^2}
  \left(1-\frac{\ln\ln k}{2\ln k}+\cdots\right),
  \label{lb}
\end{eqnarray}
obtained by Dhar and Rajesh \cite{dr21}, considering  a lattice with dimensions
$2k\times\infty$ and $k\gg 1$. The upper bound was calculated
by Gagunashvili and Priezzhev \cite{gagu}, being expressed by the equation,

  \begin{eqnarray}
    s(k)\leq\frac{\ln (\gamma k)}{k^2},
    \label{ub}
  \end{eqnarray}
  where $\gamma=\exp(4G/\pi)/2$, with $G$ being the already mentioned
  Catalan's constant. Notice that this upper limit coincides with the exact
  value of the dimer entropy on the square lattice when $k=2$.

  We can also observe that as $k$ grows, the behavior for $s(k)$
  has a tendency to approach that one predicted by Dhar and Rajesh \cite{dr21},
  $s=\frac{\ln k}{k^2}$, for the case of very large chains. Actually,
  beyond $k=5$ the difference between our values and the asymptotic
  prediction differ less than $3\%$.

\subsection{Helical boundary conditions}
\begin{figure}[h!]
\begin{center}
  \includegraphics[scale=0.75]{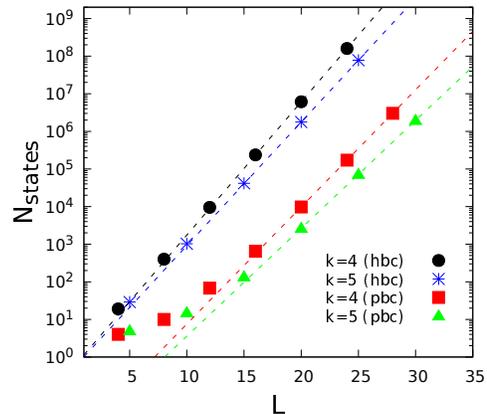}
  \caption{Dimension for the transfer matrix as a function of the
    width $L$, considering results coming from the periodical (pbc)
    and helical (hbc) boundary conditions obtained for $k=4$ and $k=5$.}
    \label{fig5}
\end{center}
\end{figure}

The transfer matrices obtained through this approach display a larger number
of states than those obtained considering periodical boundary conditions.
A comparison between those numbers can be seen in Fig. \ref{fig5}, where besides
noting the exponential dependence between the number of states and the
width $L$, for a given value of $k$, already seen in the
pbc case, we also can perceive that these numbers can be almost $1000$ times
bigger when the matrix is calculated considering helical boundary conditions.
In part this drawback is compensated by the fact that for helical boundary
conditions the transfer matrix is much sparser when compared to the case of
periodic boundary conditions, as already mentioned, but this also has the
effect that the number of iterations needed in the Power Method to reach a
selected convergence will be larger for helical boundary conditions.

Evidently, because of that, the largest value of $L$ attained for each size
of the chains is smaller than those reached for the pbc calculations. Then,
once the eigenvalues, as it also happens for the periodical case, are arranged
in sets of sizes sharing the same remainder for the division $L/k$, the
number of
elements for each set is smaller when compared with those shown in the table
\ref{tab1}. For this boundary condition those numbers are presented in
table \ref{tab3}.

\begin{table}[h!]
\begin{center}
  \begin{tabular}{|c||c|c|c|c|c|c|c|c|c|c|}
    \hline
    k&$R_0$&$R_1$&$R_2$&$R_3$&$R_4$&$R_5$&$R_6$&$R_7$& $R_8$& $R_9$\\
    \hline
    \hline
    2&14&14& & & & & & & & \\
    \hline
    3& 8 &8&7& & & & & & & \\
    \hline
    4& 6 &6&5&5& & & & & & \\
    \hline
    5& 5 &5&5&5&4& & & & & \\
    \hline
    6&4&4&4&4&4&4& & & & \\
    \hline
    7&4&3&3&3&3&3&3& & & \\
    \hline
    8&4&4&4&3&3&3&3&3& & \\
    \hline
    9&4&3&3&3&3&3&3&3&3& \\
    \hline
    10&3&3&3&3&3&3&3& & & \\
    \hline
  \end{tabular}
  \caption{Dimension of each set related to the remainder R, as it shown
    in table \ref{tab1}, but obtained for helical boundary condition.}
    \label{tab3}
\end{center}
\end{table}

\begin{figure}[h!]
\begin{center}
  \includegraphics[scale=0.4]{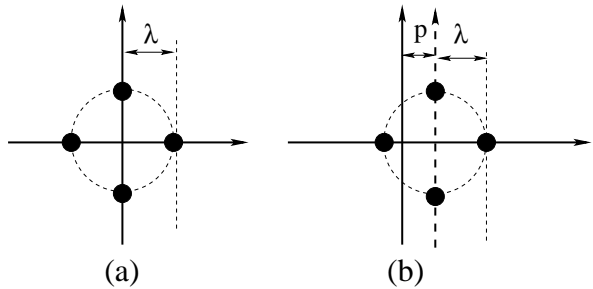}
  \caption{Illustration on how the largest eigenvalues for a transfer
    matrix $\mathcal{T}$ are distributed with the same modulus on the complex
    plane (a) and how the translation produced by the parameter $p$ turns
    the eigenvalue located at
    the real axis into the only dominant eigenvalue for a transformed matrix
    $\mathcal{T'}$. This example corresponds to the situation we get for
    $k=4$ with hbc, where the largest eigenvalue is fourfold degenerate.}
    \label{fig6}
\end{center}
\end{figure}

Another similarity found in those two approaches is the disposition for the
leading eigenvalues of the transfer matrix. Just as it happens for
the periodical
boundary conditions, the largest eigenvalue is degenerate on the
complex plane, at least one of them being located at the real axis.
Again, because
in this case the transfer matrix is even sparser than those obtained for
the periodical boundary conditions, we have used the Power Method
in order to get
this leading eigenvalue. As already mentioned in the previous discussion for
the pbc case, to circumvent this degeneracy, which puts the Power
Method in jeopardy,
we diagonalize a transformed matrix $\mathcal{T'}$, translating all the diagonal
elements from the original matrix, $\mathcal{T}$, by a real number $p$ - as
it is illustrated
by Fig. \ref{fig6}. Doing so, we
produce another leading eigenvalue free from any degeneracy and we can recover
the value we are looking for only subtracting $p$ from the largest eigenvalue
of $\mathcal{T'}$.

However, unlike the pbc case, the choice of $p$ in this situation can be a
sensitive issue. Also, we noticed that the estimates for the leading
eigenvalue as the iterations are done show a pattern which has
oscillations of a period of about $2kL$ with slowly decreasing
amplitude. This behavior is distinct to what happens for pbc, where the
convergence, after a short transient, is usually monotonical. Therefore,
great care has to be used in establishing the condition for numerical
convergence.

\begin{table}[h!]
\begin{center}
  \begin{tabular}{|c|c|c|c|c|c|}
    \hline
    $k$&$s_i(\sigma_i)$& $k$& $s_i(\sigma_i)$&$k$& $s_i(\sigma_i)$\\
    \hline
    2& 0.2915609040290(83)&3 &0.1585113(35)  &4&0.1007852(60)\\
     & 0.2915609040300(11)&  &0.15852375(20) & &0.10078178(76) \\
     &                    &  &0.15856889(35) & &0.100737621(73)\\
     &                    &  &               & &0.10066(80)\\
    \hline
   5&0.07035(16)          &6 &0.05224(14) &7&0.04045(50)\\
    &0.0704471(21)        &  &0.05209(24) & &0.04037(17)\\
    &0.07028(40)          &  &0.05247(51) & &0.04128(39)\\
    &0.07026(93)          &  &0.05297(62) & &0.09952(50)\\
    &0.0709(81)           &  &0.05321(41) & &0.06014(52)\\
    &                     &  &0.0646(79)  & &0.03978(49)\\
    &                     &  &            & &0.04075(55)\\
    \hline
    8&0.03240(44)       &9 &0.02660(40)   &10&0.02229(36)\\
     &0.03233(13)       &  &0.02645(11)   &  &0.022150(68)\\
     &0.03207(15)       &  &0.02634(64)   &  &0.022063(12)\\
     &0.03199(14)       &  &0.02627(34)   &  &0.0219984(15)\\
     &0.03193(18)       &  &0.026223(45)  &  &0.0219569(27)\\
     &0.03188(24)       &  &0.026186(82)  &  &0.0219318(15)\\
     &0.03183(30)       &  &0.02615(13)   &  &0.021913(16)\\
     &0.03204(34)       &  &0.02612(17)   &  &0.021893(55)\\
     &                  &  &0.02646(20)   &  &0.021873(90)\\
     &                  &  &              &  &0.021998(61)\\
    \hline
  \end{tabular}
    \caption{Same results shown in table \ref{tab2} for the hbc case.}
    \label{tab4}
\end{center}
\end{table}

\begin{figure}[h!]
  \begin{center}
    \includegraphics[scale=0.75]{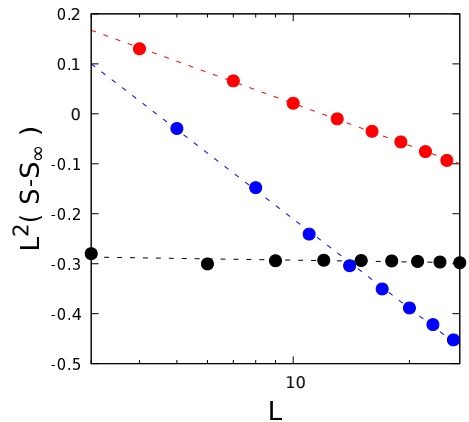}
    \caption{Behavior of the trimer entropies, separating the behavior
      of remainders $R=0,1$ and $R=2$ (black, blue and red dots, respectively)
      as a function of $\ln L$. The linear behavior for $R=1$ and $R=2$
      shows that $s(L)$ follows the behavior predicted by Eq.
      \ref{eqnew}, while the logarithmic term is absent for $R=0$,
      which follows the same dependence observed for the pbc case.
      The estimates of $s_{\infty}$ were obtained via BST extrapolation
      as discussed below.}
    \label{fig7}
  \end{center}
\end{figure}

The asymptotic behavior for the entropy, as a function of $L$, in this
case, is not the same the one found in the pbc case (Eq. \ref{eq3})
\cite{d22}. In fact,
we have found that aside from the dimer case, where the behavior is the
same predicted by Eq. \ref{eq3}, the entropies exhibit a logarithmic
correction in the form,
\begin{eqnarray}
  s(L)=s_{\infty}+\frac{1}{L^2}(A\ln L+B).
  \label{eqnew}
\end{eqnarray}
The figure \ref{fig7} shows this kind of behavior
obtained for $k=3$, where we can see clearly an evidence of such
logarithmic correction, which is not present for the pbc calculations.
This additional term shows rather clearly in our data for small values
of $k$ larger than 2, but, as expected due to the fact that as $k$ grows
there are more sets to extrapolate for each $k$ with less points in
each of them, evidence is not so clear for larger rods.

Despite that, we still are able to perform the extrapolations and
determine the asymptotic value for the entropy, $s_{\infty}$, using
the same BST approach.
Unlike, what is discussed in the appendix of \cite{henkel}, where
a function with a logarithmic correction is analyzed showing a
poor performance of the method, our calculations, despite displaying
some fluctuations for the average value and its uncertainty, were less
affected by the presence of the logarithm term. However, while for pbc
we have applied the extrapolation considering a fixed value for the
parameter $\omega$, which was set at $\omega=2$, in the hbc case,
we estimate the value of $s_{\infty}$ considering two cases: for
the sets of remainder $R=0$ we have kept $\omega=2$, just as we have
done for the ccp case. This was motivated by the fact that, at least for
rather small values of $k$, we found evidences that $A=0$ in this case.
However for sets associated with
non-vanishing remainders, we have determined the entropy as a function of
$\omega$ over a domain $\omega\in[1,2]$, adopting the value of $s_{\infty}$,
which minimizes the uncertainty defined by Eq. \ref{eq5}.

Doing so, we obtain the results shown in table \ref{tab4}, where we notice
that, in some cases, the uncertainties have the same order of magnitude as
those found from pbc calculations, although they were obtained from
smaller sets of values of the entropy (see table \ref{tab3}). This is
possible because even with less elements to use in the extrapolations,
the values for the entropies, in the hbc case, are closer to the
asymptotic limit when compared to those obtained with pbc. While for
the periodical boundary conditions, as we can see from Fig. \ref{fig2}(a),
this relaxation can be quite slow, particularly for the values in
the set with remainder $R=0$, the same does not occur for helical
boundary conditions. Then, if this boundary condition is somehow
handicapped by smaller values of $L$ attained, the extrapolation is
not much affected, since the values for finite widths
are closer to their asymptotic limit.

Given the results shown in table \ref{tab4} we proceed to the final values
for the entropy per site, considering rigid chains of size $k$ placed at
the sites of a square lattice, using the Eq. \ref{eq12}. The final results
for such entropies are shown in table \ref{tab5}. We do not have a complete
agreement between the results obtained from the ccp and hbc extrapolations.
We notice that for six values of $k$ (3, 4, 5, 8, 9, and 10) the confidence
intervals for the entropy per site obtained for periodic and helical boundary
conditions have no intersection, while for dimers ($k=2$) both intervals
contain the exact result and for trimers they are also consistent, although
more precise, with the previous result by Ghosh, Dhar, and Jacobsen
\cite{ghosh}. We suspect that the inconsistencies are due to a possible
underestimation of the uncertainties by the BST method due to the
logarithmic term in the asymptotic behavior of the approach to the
two-dimensional limit of the entropy per site for helical boundary conditions
when the remainder is non-vanishing, together with the fact that as $k$ grows
the number of points to extrapolate in each case becomes smaller. The
relative discrepancies increase with $k$, which may be an indication of
the effect of the reduction of the number of cases to extrapolate on the
quality of the results. We thus believe that in general the estimates
provided by the results from strips with periodic boundary conditions
are more reliable. In general, it is apparent that the estimates obtained
here are essentially consistent with the ones obtained through computer
simulations \cite{pasinetti}, but are more accurate.

Considering the essential divergence between the estimates obtained
for both boundary
conditions using the BST method (the gap between the confidence intervals
for $L=10$ is of the order of 30 times the uncertainty), we have also produced
estimates for helical boundary conditions using a simple alternative method:
for each size $k$ of the rods, we find the values of $s_\infty$ $A$ and $B$ of
the asymptotic behavior Eq. \ref{eqnew} which reproduce the three
entropies for the largest widths $L$ for each remainder $R=0,1,2,\ldots,k-1$.
This procedure thus
produces $k$ estimates for $s_\infty$, besides estimates for the amplitudes
$A$ and $B$. The mean value of the estimates for $s_\infty$ and its
dispersion for each rod length $k$ are displayed in table \ref{tab6}. Of
course this procedure is rather crude, since to estimate the
entropies of the strips with three different widths, we use the same
weight for different values of $L$, but we notice that the results
are compatible with the exact value
and the extrapolated estimates for dimers, with the estimate in \cite{ghosh}
and the estimate for periodic boundary conditions for trimers. For
$k=4$, 5, 6, and 7 we also have an agreement with the estimates coming
from the results with pbc, but for the rest the confidence intervals
for the new estimates are below the ones obtained with pbc. So we see
that, although larger uncertainties are obtained in the alternative
extrapolation procedure for hbc, there are still some cases where
the results are not compatible with the ones provided by the
calculations with pbc, although the relative discrepancies are much
lower than the ones found comparing results for both boundary
conditions using the BST extrapolation procedure and these cases are
restricted to the larger values of $k$, where it is clear that the method
in all approaches leads to less precise results.

\begin{table}[h!]
\begin{center}
  \begin{tabular}{ccccccc}
    \hline
    \hline
    $k$& & TM (pbc)& & TM (hbc) & & MC
    \cite{pasinetti}\\
    \hline
    2& &0.29156090404(14)  & & 0.2915609040293(66) & & 0.2930(20)\\
    3& &0.15850494(19)     & & 0.15853458(17) & &0.1590(20)\\
    4& &0.1007670(36)      & & 0.100738034(73) & &0.1010(20)\\
    5& &0.07038320(58)     & & 0.0704470(21) & &0.0700(30)\\
    6& &0.0522274(65)      & & 0.05232(11) & &0.0520(30)\\
    7& &0.0404963(64)      & & 0.04048(14) & &0.0400(30)\\
    8& &0.0324516(39)      & & 0.032088(67)& &0.0320(30)\\
    9& &0.0267234(85)      & & 0.026266(23)& &0.0270(30)\\
    10&& 0.022337(12)      & & 0.0219643(10)& &0.0210(30)\\
    \hline
    \hline
  \end{tabular}
  \caption{Results for the entropy of $k$-mers in the full lattice limit.
    The first column contains averages calculated
    from the entropies shown in table \ref{tab2} using the relations shown
    in Eq. \ref{eq12}, for periodic boundary conditions. The second column
    shows averages calculated from the entropies shown in table \ref{tab4},
    for helical boundary conditions. The third column displays the results
    obtained by Pasinetti {\it{et. al}} through computer simulations
    \cite{pasinetti}.}
    \label{tab5}
\end{center}
\end{table}

\begin{table}[h!]
\begin{center}
  \begin{tabular}{ccc}
    \hline
    \hline
    $k$& & TM (hbc) \\
    \hline
    2& & 0.29156104(25) \\
    3& & 0.158511(14) \\
    4& & 0.100822(86) \\
    5& & 0.070381(70) \\
    6& & 0.052213(58) \\
    7& & 0.04029(21) \\
    8& & 0.03210(15) \\
    9& & 0.02632(13) \\
    10&& 0.02202(11)  \\
    \hline
    \hline
  \end{tabular}
  \caption{Results for the entropy of $k$-mers in the full lattice limit.
    Estimates obtained using an alternative procedure to extrapolate
    the results for helical boundary conditions, described in the text.}
    \label{tab6}
\end{center}
\end{table}

\section{Final Discussion and Conclusion}
\label{conc}
In the present work, we have dealt with the problem of determining
the configurational entropy for colinear chains of size $k$, named $k$-mers,
fully covering a square lattice. To do so, we have employed transfer
matrix calculations using three different constructions. Two of them
were employed for periodical boundary conditions, the so-called {\it{Usual
Approach}}, already used by Ghosh {\it{et. al}} \cite{ghosh} to obtain
the entropy for trimers ($k=3$), and the {\it{Profile Method}}, based
on the calculation developed by Dhar and Rajesh \cite{dr21} in order to
estimate a lower boundary for the value of the entropy as a function
of the chain size, $k$, considering $k\gg 1$. To our knowledge, this second
approach was never used in the transfer matrix method and it has been
useful to deal with this problem. Since we seek to determine
the entropy for full coverage in the thermodynamic limit from the results
obtained for the entropy of the $k$-mers placed on strips with 
finite widths equal to $L$, our results tend to be better when we reach
large values of $L$. The {\it{Profile Method}}, in the majority of cases,
produces transfer matrices with smaller dimensions than those obtained
via the {\it{Usual Approach}}, allowing us to obtain better numerical results
for the entropies. We notice that in the {\it Usual Approach} the
entropy is directly related to the leading eigenvalue of the transfer matrix,
while in the {\it Profile Method}, which is grand-canonical, it is necessary
to find the value of the activity of a $k$-mer which corresponds to a leading
eigenvalue with a unitary modulus. So, while in the first approach
we need to find the leading eigenvalue only once, in the second approach it
is necessary to repeat this operation several times to reach the required
numerical precision. Nevertheless, the {\it Profile Method} allowed us to reach
larger widths. Another construction we have applied for these calculations
was the {\it{Usual Approach}} considering helical boundary conditions. However,
even being less effective to reach large values of $L$, this approach
has a tendency to generate values closer to the asymptotic limits associated
with the thermodynamic limit, although displaying greater uncertainties.

Although we have not presented results on details about the convergence of
the results of the entropies on strips of finite widths to the two-dimensional
values, as was, for instance, done for trimers in \cite{ghosh} it was clear that
the scaling form Eq. \ref{eq3} is followed by our results, for 
periodic boundary conditions. This is an indication that the
phase in the full lattice limit is critical and conformal invariant for
periodic boundary conditions. We plan to come back to this point in the future.

Our results show values that are in accordance with some previous results
in the available literature, such as the case for dimers ($k=2$), the only
case which was exactly solved and for which our result agrees up to the 11th
decimal place, and also the for trimer case, where the entropy
obtained here agrees
with the one estimated by Ghosh {\it{et. al}} in \cite{ghosh}. Another source
for comparison are the simulational results obtained by Pasinetti
{\it{et. al}} \cite{pasinetti} which also are in complete agreement with our
values, although they are less precise. We may also compare our results
with recent estimates for the entropies for the same problem provided by a
sequence of Husimi lattice closed form approximations \cite{rso22}, which
are numerically exact solutions on treelike lattices that may be considered
{\it beyond mean field} approximations. These results, for $k$ in the
range $\{2-6\}$, in a similar way to
ours, become less precise for growing values of $k$. While the relative
differences between the present
and the former estimates are of the order of 3 \% for $k=2,3$, they reach
about 40 \% for the higher values of $k$. It is also noteworthy that the
behavior displayed by the entropies $s$ and the sizes $k$ seemingly obey
the relation predicted by Dhar and Rajesh \cite{dr21}, $s\approx\ln k/k^2$,
when $k\to\infty$. As it has been mentioned previously, from $k=5$ up to $k=10$
our results differ from that expression by less than 3$\%$.

\section{Acknowledgments}
This work used computational resources of the ``Centro Nacional de
Processamento de Alto Desempenho" in S\~{a}o Paulo (CENAPAD-SP, FAPESP).
We also acknowledge the help by Rogerio Menezes for his aid with some
other computational resources used in our calculations. We thank Prof. Deepak
Dhar for a critical reading of the manuscript and for pointing out to us
the logarithmic correction in the asymptotic behavior of the results for
helical boundary conditions.


\begin{thebibliography}{99}
\bibitem{d22}This point was communicated to us by Prof. Deepak Dhar.
\end{thebibliography}
\end{document}